\documentclass[structabstract]{aa}

\usepackage{graphicx}
\usepackage{subfigure} 
\usepackage{float}
\usepackage{psfrag}
\usepackage{amsmath}
\usepackage{amssymb}
\usepackage{ulem}
\usepackage{natbib} 
\usepackage{txfonts}
\usepackage{lscape}

\begin{document}
   \title{Multi-wavelength observations of afterglow of GRB\,080319B and the modeling constraints.
\thanks{Based on observations obtained with
the 0.22\,m telescope at Russia
the 0.7\,m telescope at of Kharkov University, Ukraine,
the 0.8\,m telescope at Observatorio del Teide (IAC-80), Spain
the 1.2\,m Mercator telescope at La Palma, Spain,
the 1.5\,m telescope of Maidanak observatory Uzbekistan,
the 2.0\,m IGO Telescope at IUCAA Pune, India,
the 2.5\,m NOT,
the PdB millimeter interferometric array France,
the RATAN-600 Radio Telescope at Russia and
the RT-22 radio telescope of CrAO, Ukraine.
}}

\titlerunning{
Multi-wavelength observations of afterglow....  }

   \author{S. B. Pandey\inst{1}
          \and
           A. J. Castro-Tirado\inst{2}, M. Jel\'{\i}nek\inst{2}
          \and
          Atish P. Kamble\inst{3}
          \and
          J. Gorosabel\inst{2}
          \and
          A. de Ugarte Postigo\inst{4}
          \and
          S. Prins\inst{5}, R. Oreiro\inst{5}
          \and
          V. Chantry\inst{6}
          \and
          S. Trushkin\inst{7}
          \and
          M. Bremer\inst{8}, J. M. Winters\inst{8}
          \and
          A. Pozanenko\inst{9}
          \and
          Yu. Krugly\inst{14},  I. Slyusarev\inst{14}
          \and
          G. Kornienko\inst{10}, A. Erofeeva\inst{10}
          \and
          K. Misra\inst{11}, A. N. Ramprakash\inst{11}, V. Mohan\inst{11}, D. Bhattacharya\inst{11}
          \and
          A. Volnova\inst{12}
          \and
          J. Pl\'{a}\inst{13}
          \and
          M. Ibrahimov\inst{15}
          \and
          M. Im\inst{16}
          \and
          A. Volvach\inst{17}
          \and
          R. A. M. J. Wijers\inst{3}
          }

  \institute{
          Aryabhatta Research Institute of Observational Sciences (ARIES), Manora Peak, Nainital, India, 263129.
           \and
          Instituto de Astrof\'{\i}sica de Andaluc\'{\i}a (IAA-CSIC),
          Apartado de Correos, 3.004, E-18.080 Granada, Spain.
           \and
          Astronomical Institute ``Anton Pannekoek'', Kruislaan, 403, 1098SJ, Amsterdam, The Netherlands.
           \and
          European Southern Observatory, Casilla 19001, Santiago 19, Chile.
           \and
          Institute of Astronomy, Katholieke Universiteit Leuven, Celestijnenlaan 200D, 3001 Leuven, Belgium
           \and
          Institut d\'Astrophysique et de Geophysique, Universite de Liege, Allee du 6 Aout 17, Sart
          Tilman (Bat. B5C), Liege 1, Belgium.
           \and
          Special Astrophysical Observatory of RAS, Nizhnij Arkhyz, Karachai-Cherkessia, 369167 Russia.
           \and
         Institute de Radioastronomie Millim\'etrique (IRAM), 300 rue de la Piscine, 38406 Saint Martin d\'\rm H\'eres, France.
           \and
         Space Research Institute of RAS, Profsoyuznaya, 84/32, Moscow 117997, Russia.
           \and
         Ussuriisk Astrophysical observatory, Far East Branch of RAS, Ussuriisk Region, Gornotaejnoe, 692533, Russia.
           \and
         Inter-University Center for Astronomy and Astrophysics (IUCAA), Pune, Ganeshkhind, Post-bag No. 4, India.
           \and
        Sternberg Astronomical Institute, Moscow State University, Universitetsky
        pr., 13, Moscow 119992, Russia.
           \and
        Instituto de Astrof\'{\i}sica de Canarias (IAC), Via L\'actea s/n,
   E-38205 La Laguna (Tenerife), Spain.
           \and
        Astronomical Institute of Kharkov National University, 35 Sumskaya, Kharkov, 61022, Ukraine.
           \and
        Ulugh Beg Astronomical Institute, Tashkent 700052, Uzbekistan.
           \and
        Dept. of Physics \& Astronomy, Seoul National University, 56-1 San,
        Shillim-dong, Kwanak-gu, Seoul, Korea.
           \and
        SRI Crimean Astrophysical Observatory, Nauchny, Crimea, 98409, Ukraine.
          }


 \abstract{}{}{}{}{}


  \abstract
    {We present observations of the afterglow of GRB\, 080319B at optical, mm and
    radio frequencies from a few hours to 67 days after the burst.}
   {To understand the nature of this brightest explosion based on the observed properties
    and it's comparison with the afterglow models.}
   {Present observations along with other published multi-wavelength data have been used
    to study the light-curves and spectral energy distributions of the burst afterglow.}
   {Our results show that the observed features of the afterglow fits equally good with 
   the Inter Stellar Matter and the Stellar Wind density profiles of the circum-burst medium. In case of
    both density profiles, location of the maximum synchrotron frequency $\nu_m$ is below
    optical and the value of cooling break frequency $\nu_c$ is below $X-$rays, $\sim 10^{4}$
    s after the burst. Also, the derived value of the Lorentz factor at the time of naked eye
    brightness is $\sim 300$ with the corresponding blast wave size of $\sim 10^{18}$ cm.}
   {The numerical fit to the multi-wavelength afterglow data constraints the values of
    physical parameters and the emission mechanism of the burst.}


   \offprints{ \hbox{e-mail:\,{\tt shashi@aries.res.in}}}

      \maketitle
%

\section{Introduction}

The very early time observations of Gamma-ray bursts (GRBs) during the prompt
emission or during the early afterglow phase have been one of the major 
contributions during the {\it Swift} era (Zhang 2007). These observations play 
key role to uncover the physical mechanisms underlying these energetic 
cosmic explosions (Me\'sza\'ros 2006). The early time multi-wavelength observations are also
very useful to constrain the afterglow models hence the nature of the possible 
progenitors and the ambient media surrounding the GRBs (Castro-Tirado et al. 1999; Piran 1999). 
Among the well-known examples of early observations in the pre-{\it Swift} era are
GRB\, 990123 (Akerlof et al. 1999); GRB\, 041219 (Vestrand et al. 2005) whereas
GRB\, 050820A (Vestrand et al. 2006); GRB\, 060111B (Klotz et al. 2006); GRB\, 060210
(Stanek et al. 2007), and GRB\, 071010B (Wang et al. 2008) are good examples in the 
post-{\it Swift} era. The statistics of such examples of long-duration 
GRBs with very early time afterglow observations have been improved because of the 
precise on-board localization by {\it Swift} and co-ordinated observations with the ground 
based robotic optical telescopes (Gehrels et al. 2004, Gomboc et al. 2006). 
Out of these known examples the observed optical and $\gamma$-ray prompt emission 
is contemporaneous for GRB\, 041219 and GRB\, 050820A whereas for GRB\, 990123 and 
GRB\, 050401 the peak of the prompt optical observations have been recorded after 
the $\gamma$-ray emission phase. 

According to the standard relativistic ``Fireball'' model (Rees \& Me\'sza\'ros 1992; 
Me\'sza\'ros \& Rees 1997; Panaitescu et al. 1998), the GRB prompt emission is due
to the internal shocks (Narayan et al. 1992; Rees \& Me\'sza\'ros 1994; Sari \& Piran 1997a,b). 
The observed steep decay in the early afterglow emission, that lies between the prompt 
emission and the afterglow, are also explained in terms of ``high latitude emission'' 
(Nousek et al. 2006) in the case of {\it Swift} GRBs, irrespective of the type of shock and the 
radiative process involved. The very early time optical observations of several GRB 
afterglows are explained in terms of reverse shock (Kobayashi 2000) or/and forward shock 
(Rees \& Me\'sza\'ros 1992; Katz 1994) origin. The overall afterglow behavior of long-duration 
GRBs including the above mentioned examples are explained in terms of 
``collapse of very massive stars'' i.e.  ``Collapsars'' as the most favored 
progenitor (MacFadyen \& Woosley 1999). By now the majority of long-duration GRB afterglows 
have been explained in terms of constant ambient density i.e. Inter Stellar Medium 
({\it ISM}, $\rho \propto r^{0}$) models (Sari, Piran, Narayan 1998; Wijers \& Galama 1999; 
Panaitescu \& Kumar 2002), although stellar wind medium ({\it WM}, $\rho \propto r^{-2}$) profile 
(Chevalier \& Lee 2000a,b; Li \& Chevalier 2001) are the natural consequences for the 
massive star environments (Zhang 2007), where $\rho$ and $r$
are the ambient density and the distance from the center of the progenitor star respectively. 
The value of ambient density is constrained by the parameters ``number density'' $n$ and the 
``wind parameter'' $A_{\ast}$, respectively for the {\it ISM} and {\it WM} models. 
Also, there are certain cases of long-duration GRB afterglows which have been explained in 
terms of both density profiles in form of the transition from {\it WM} density profile at 
early times to {\it ISM} density profile at later epochs of observations, e.g.
GRB\, 050904 (Gendre et al. 2007) and GRB\, 050319 (Kamble et al. 2007). 

GRB\, 080319B was triggered (trigger = 306757) by {\it Swift}-BAT (15-350 keV) at 
$T_0$=06:12:49 UT on March 19, 2008 (Racusin et al. 2008a) and was simultaneously detected 
by the Konus-Wind (20 keV-15 MeV) satellite (Golenetskii et al. 2008). The optical emission 
were started 2.75$\pm$5 s after the BAT trigger, captured by the wide field robotic telescope 
``Pi of the Sky'' (Cwiok et al. 2007, 2008) and also by TORTORA (Molinari et al. 2006, 
Racusin 2008b), Raptor-Q (Wozniak et al. 2009) at later epochs. 
For a few seconds after the burst, the observed 
prompt optical flash of GRB\, 080319B was visible even for the unaided eye in dark skies 
(peaked at a visual magnitude of 5.3 around 18 s after the onset of the burst, as observed 
by TORTORA), breaking the record of a handful of known cases in which bright optical{\it-NIR} 
prompt emissions have been observed (Jelinek et al. 2007). 
{\it Swift}-UVOT slewed towards the burst 51 s after the trigger (Holland \& Racusin 2008) 
and later many other ground and space based multi-wavelength facilities joined in, 
as summarized in detail by Bloom et al. (2009), Racusin et al. (2008b) and references therein.

The measured redshift value of the burst ($z=0.937$, Vreeswijk et al. 2008) corresponds to a
luminosity distance of $d_L=6.01\times10^{3}$ Mpc (with the cosmological parameters $H_0=71~\rm
km~s^{-1}Mpc^{-1}$, $\Omega_{M}=0.27$, $\Omega_{\Lambda}=0.73$) and an equivalent isotropic energy in
$\gamma$-rays $E_{\gamma,iso}=1.4\times10^{54}\rm $ erg ($20\,{\rm keV}-7\,{\rm MeV}$), 
which is among the highest ever measured for GRBs. The inferred high luminosity of the burst
after correcting to the  galactic and host extinction in the burst direction (Wozniak et 
al. 2009) implies that such an event could have been easily detected at redshift $z = 17$ with 
meter-class telescopes in the $K-$band (Bloom et al. 2009, Sagar 2005). Also, the observed value 
of the flux density of the optical flash of GRB\, 080319B ($\sim20$ Jy) is about 4 orders of 
magnitude higher than the peak flux density at the $\gamma$-rays ($\sim14$ mJy), the highest ever 
observed for GRBs (Yu et al. 2008). The significant excess of the prompt optical flux 
in comparison to the extrapolated $\gamma$-ray spectrum indicates  
different emission components for the two observed frequencies (Racusin et al. 2008b).

Thus, it is clear that multi-wavelength afterglow observations of GRB 080319B provide an unique
opportunity to study the nature of this energetic cosmic explosion in detail. The observed spectral
and temporal coverage of the afterglow make the burst as one of the best known examples to test 
the theoretical afterglow models (Pandey et al. 2003, Resmi et al. 2005). In this paper we 
summarize the radio, Millimeter wave (mm) and the optical observations 
of the afterglow in \S 2. The afterglow properties are discussed in \S 3. The modeling of the
multi-wavelength afterglow data and the derived parameters are described in \S 4. In the last 
section we summarize our results in the context of observed and modeled parameters of the burst.

\begin{figure}[h]
   \centering
\includegraphics[width=5.5cm,angle=-90]{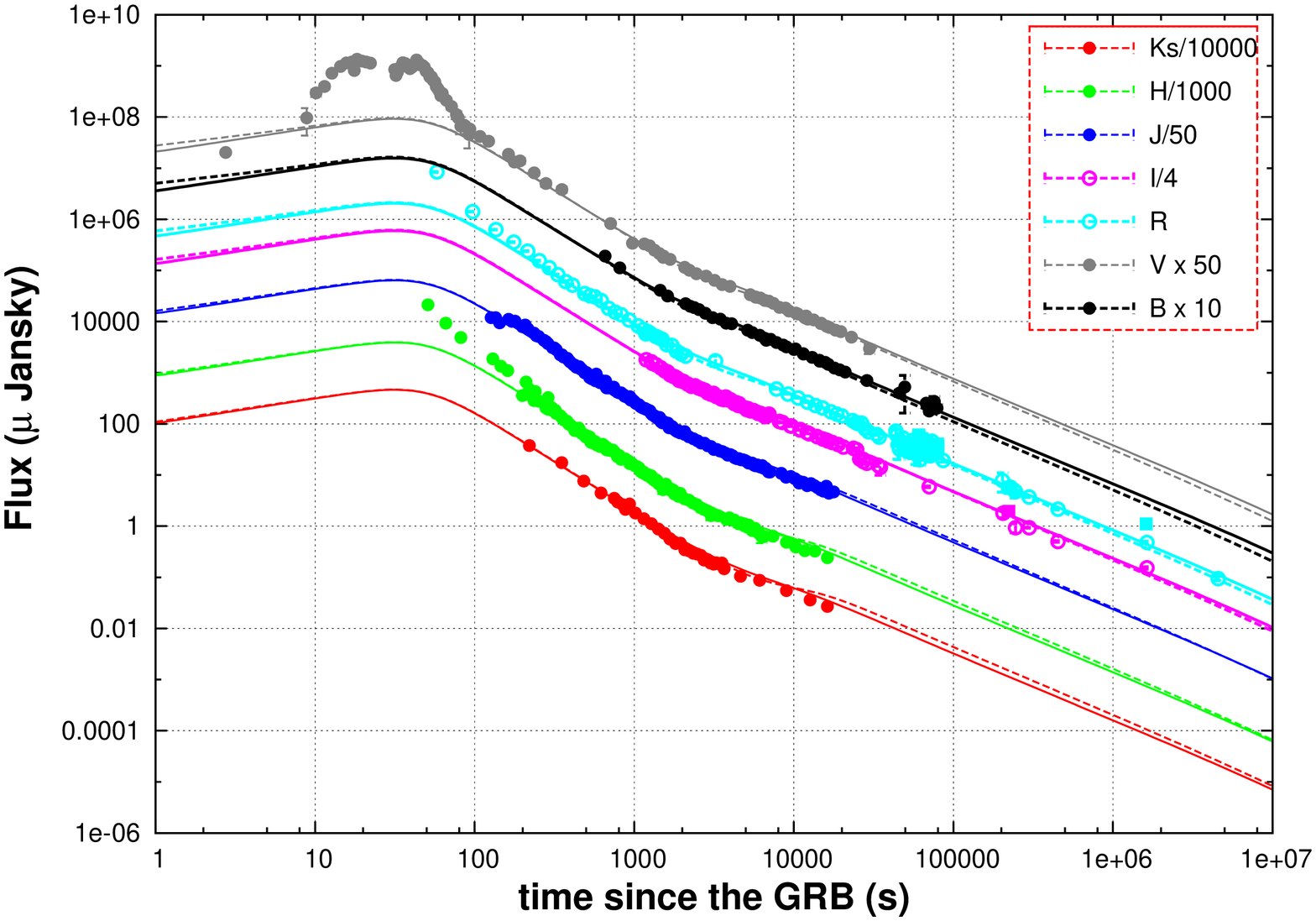}
\includegraphics[width=5.5cm,angle=-90]{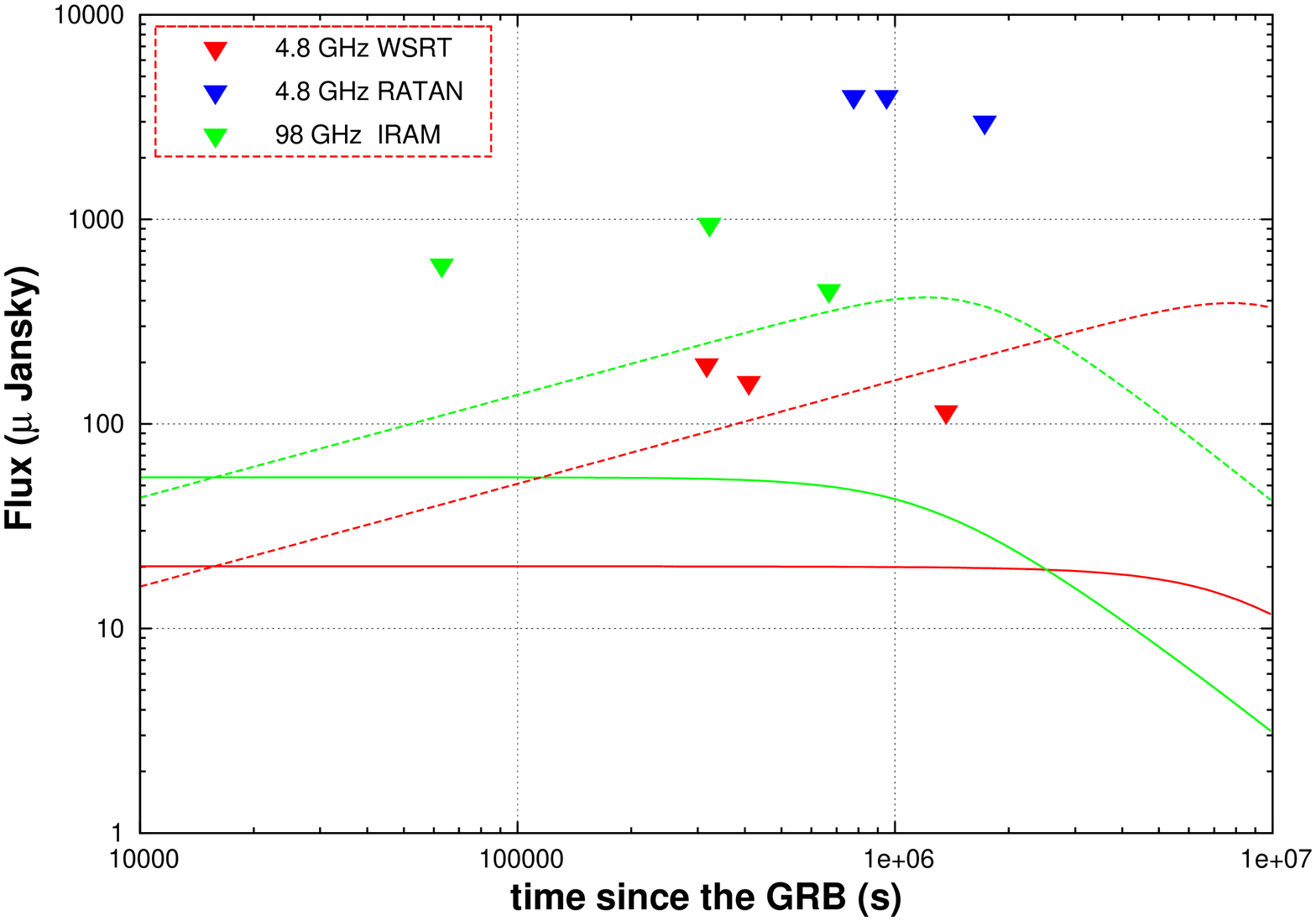}
\caption{{\bf Light curves of the afterglow of GRB\, 080319B :}
The optical{\it-NIR} and $X-$ray afterglow is shown in the top panel and the 
radio data in the bottom panel. The present optical observations are plotted along with  
other published optical and $X-$ray data, taken from the literature 
(Bloom et al. 2009, Racusin et al. 2008b). The FS-RS model is used
to fit the afterglow light curves. The solid lines in the plots
represent the resultant best fit {\it WM} model whereas dotted line
show the best fit {\it ISM} model. The radiation due to the RS 
peaks around $\sim ~50~s$ and dominates the afterglow up to a few
thousand seconds. We have considered the case of 'Thick Shell' as
discussed in Kobayashi et al.(2000) to calculate the RS afterglow.
The optical band is assumed to be between the $\nu_m^{RS}$ and $\nu_c^{RS}$ i.e.
the values of $\nu_m$ and $\nu_c$ during the reverse shock phase respectively
and the value of electron energy distribution index to be $p=2.32$. It can be
clearly seen from the plots that the optical flash corresponding
to the 'naked eye' brightness of the afterglow can not be explained
as being due the RS. The decline of the $R$ and $H$ bands brightness
before a few hundred seconds is faster than that predicted by
the RS-FS model used here. This could be the high-latitude 
emission from the optical flash. The FS radiation starts dominating
the optical light curves after about a few thousand seconds.
There is no sign of any jet break at least until about $10^{6}~s$.
The predicted model results at $X-$ray frequencies are also plotted along
with the observed $X-$ray light-curve (solid squares), showing clear deviations.
The bottom panel shows the radio-mm observations of the afterglow.
The inverted triangles represent the $5\sigma$ upper limits
on the afterglow flux and the solid lines are the best fit
estimates of the afterglow flux due to the FS. The radio afterglow
due to the RS is expected to have died down significantly by
the time the radio observations were conducted.}
\end{figure}

\begin{figure}[h]
   \centering
\includegraphics[width=5.5cm,angle=-90]{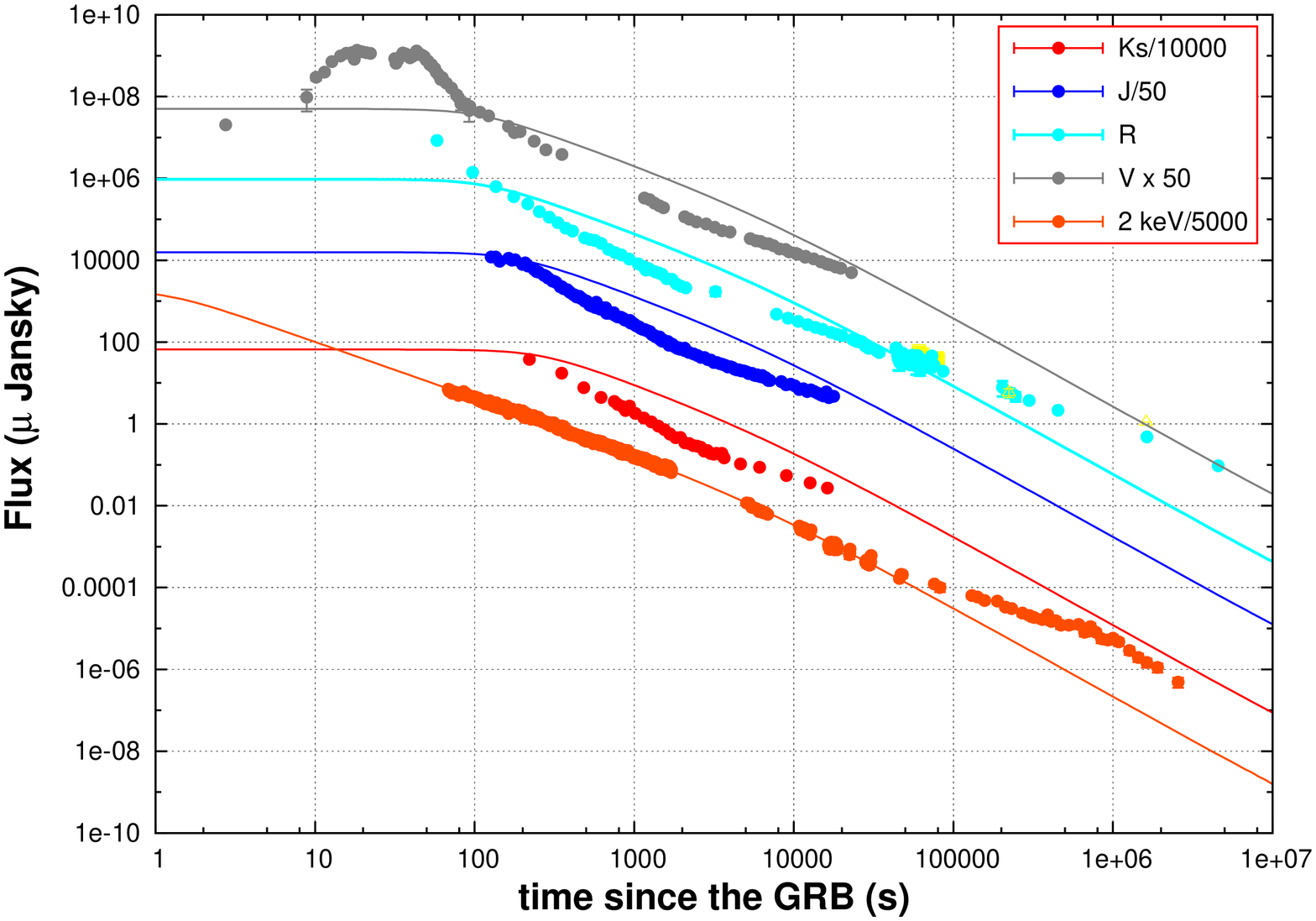}
\includegraphics[width=5.5cm,angle=-90]{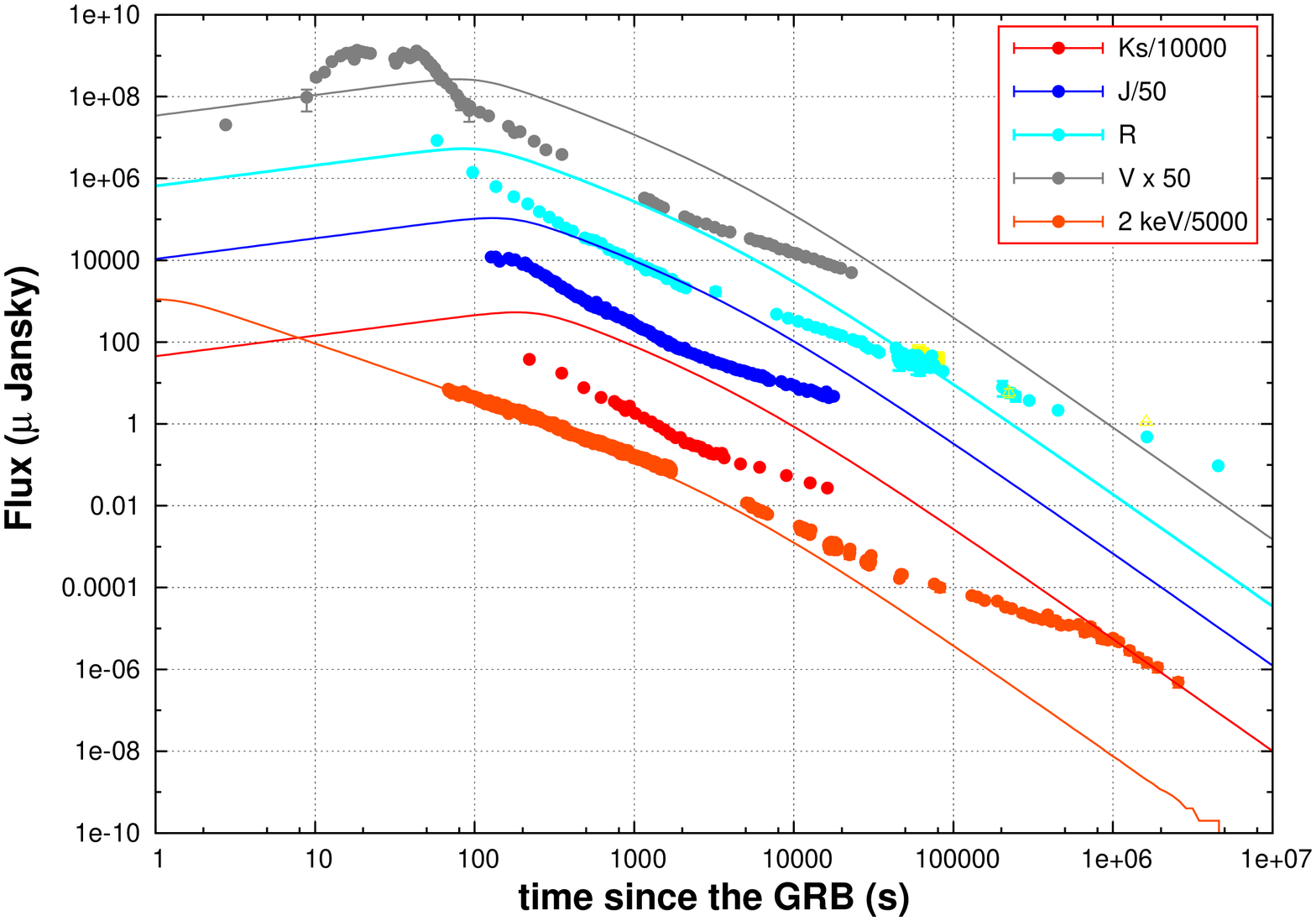}
\caption{{\bf The afterglow of GRB 080319B - a comparison with the double jet model:}
The model light-curves shown in top and bottom panels are plotted for the {\it WM} and {\it ISM} 
cases, respectively. The model here assumes the double jet scenario. The contribution to the 
afterglow only due to the narrow jet is shown in the figure. The narrow jet is assumed to be 
responsible for the the early $X-$ray afterglow. The estimated optical radiation due to the narrow 
jet is much brighter than the observed optical afterglow, which invalidates the double jet model 
(Racusin et al. 2008b) for GRB 080319B.}
\end{figure}

\section{Observations}

\subsection{Millimeter wave and Radio Observations}

Observations at millimeter frequencies of GRB\, 080319B afterglow were performed 
with the Plateau de Bure (PdB) Interferometer (Guilloteau et al. 1992) in a six-antenna 
extended configuration on the dates listed in Table 1. The afterglow was detected in 
the first round of observations starting 0.5 days after the burst and upper limits were 
established at later two epochs. Also, Radio observations were performed using RATAN-600 
at 4.8 GHz starting 9 to 19 days after the burst and the afterglow was not detected.
The afterglow was also monitored using the radio telescope RT-22/CrAO at 2.0 and 8.0 GHz 
on 25th May 2008. The upper limits (3$\sigma$) from RATAN-600 and RT-22/CrAO observations 
are also listed in Table 1. 

\begin{table}
\caption{Log of Millimeter wave and radio observations of the GRB\, 080319B afterglow.
The first IRAM data point has already been published in Racusin et al. (2008b). The PdB
millimeter interferometer array is in France, RATAN-600 Radio Telescope is situated 
close to Zelenchukskaya, Russia, and RT-22 is in Ukraine.}
\begin{center}
\scriptsize
\begin{tabular}{ccccll} \hline
Start & End & Frequency & Flux & Telescope \\
time (UT)&time (UT)&(GHz)&center (mJy) &  \\   \hline
2008 Mar 20.0& Mar 20.5&97.0&$+$0.41$\pm$0.12 & IRAM \\
2008 Mar 23.0& Mar 23.5&97.0&$+$0.35$\pm$0.19 & IRAM \\
2008 Mar 27.0& Mar 27.5&97.0&$+$0.20$\pm$0.09 & IRAM \\ \\
2008 Mar 28.0& Mar 28.5&4.8 &$<$4.0 & RATAN \\
2008 Mar 30.0& Mar 30.5&4.8 &$<$4.0 & RATAN \\
2008 Apr 08.0& Apr 08.5&4.8 &$<$3.0 & RATAN \\ \\
2008 May 25.0& May 25.5&2.0&$<$3.0 & RT-22 \\
2008 May 25.0& May 25.5&8.0&$<$3.0 & RT-22 \\
\hline
\end{tabular}
\end{center}
\end{table}

\subsection{Optical Observations}

Our optical observations were performed using several Telescopes (0.22m SR-22; 0.7m AZT-8; 
0.8m IAC; 1.2m Mercator, 1.5m Maidanak (AZT-22), 2.0m IGO, 2.5m NOT(+ALFOSC)) in Far East of 
Russia, Europe, Middle Asia and India starting from 0.5 day to 19 days after the burst. 
IRAF and DAOPHOT softwares were used to perform data reduction using standard techniques. 
The $B, V, R$ and $I$ data has been calibrated using the nearby secondary standards as given 
in Henden (2008). The Gunn $r$ data were calibrated using nearby calibrators. The log of 
our optical observations along with the details are given in Table 2.

\section{Properties of the burst}

\begin{figure*}[t]
   \centering
\hspace{2.0cm}\includegraphics[width=6.0cm,height=5.0cm,angle=-90]{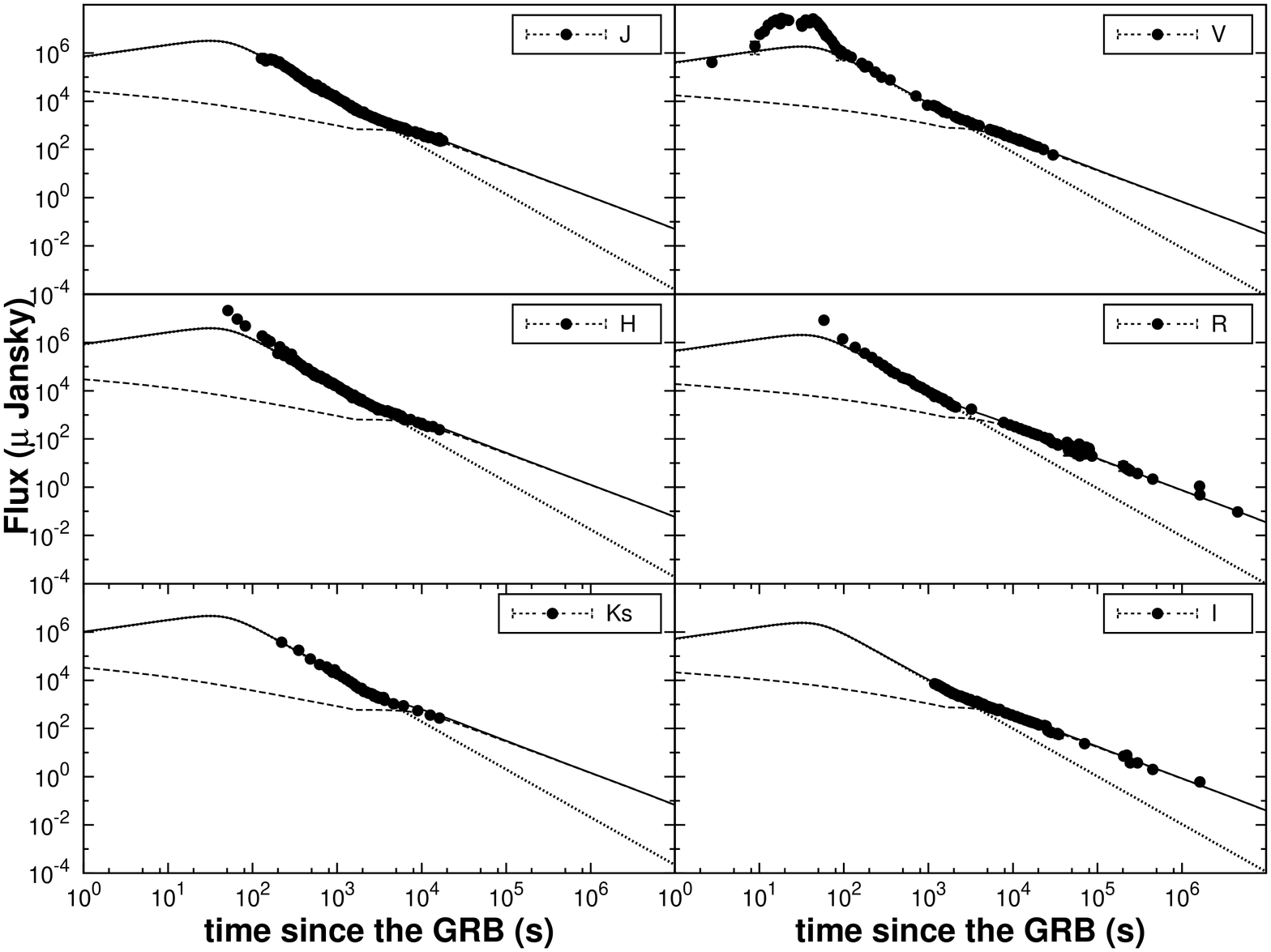}
\hspace{2.0cm}\includegraphics[width=6.0cm,height=5.0cm,angle=-90]{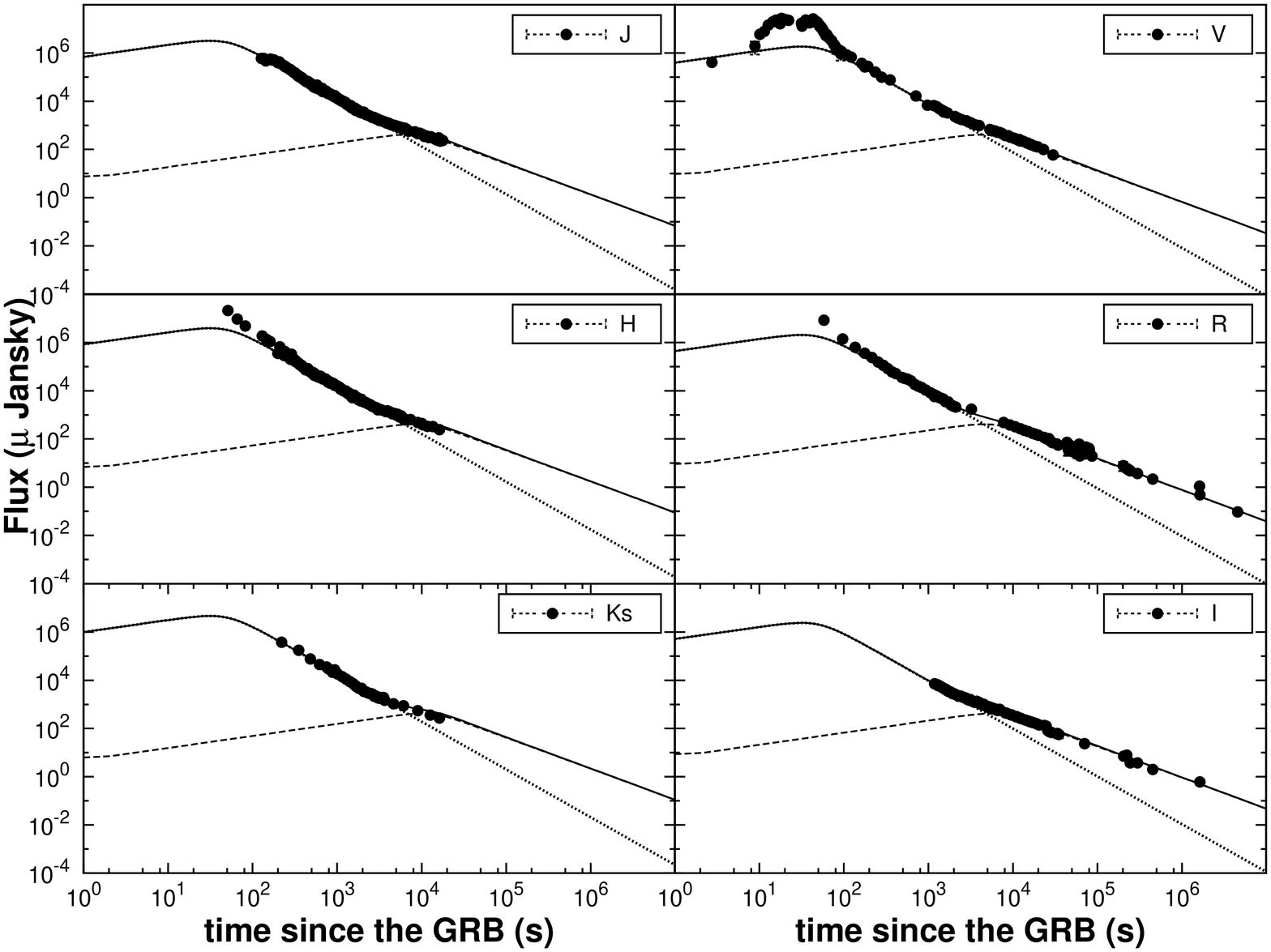}
\vspace{1.0cm}
\caption{The contributions due to the RS and FS to the afterglow of GRB 080319B at optical-$NIR$ bands 
are represented by using the dashed and dotted lines respectively. The addition of the two components is 
shown using the solid line. The left and right panels are {\it WM} and {\it ISM} models, respectively.}
\end{figure*}

\subsection{Prompt emission phase}

The prompt optical light-curve present a good correlation with the $\gamma-$ray 
light-curve, thanks to good time resolution (Racusin et al. 2008b). This correlation suggests 
that radiation at both frequencies originate from the same physical source under the 
assumptions of internal shock model (Me\'sza\'ros \& Rees 1999; Kumar \& Panaitescu 2008). 
There is no considerable evolution of the $X-$ray hardness ratio from very early to late 
times of the observations with an average photon index $\Gamma = -1.81\pm0.04$ 
(Bloom et al. 2009). 
There are no multi-band optical observations during the prompt emission phase of the burst. 
The observed spectral indices around 430s and 875s after the burst also show negligible 
evolution at optical frequencies (Bloom et al. 2009). 
The reverse shock or forward shock origin of the observed prompt optical emission is ruled out 
based on the observed values of temporal decay indices to be too steep than the expected 
one (Sari, Piran \& Narayan 1998; Kobayashi 2000). Also, the observed constancy of the optical 
pulse width with time during the prompt emission phase (Ramirez-Ruiz \& Fenimore 2000) does not
support the reverse shock origin.
The temporal coincidence between the prompt optical and $\gamma-$rays of the burst also indicate 
that they might have been originated from the same emitting region or two regions sharing the 
same dynamical behavior (Yu et al. 2008). The correlation of spectral lag evolution with the 
observed prompt optical emission of GRB\, 080319B also shows that they come from the same 
astrophysical origin and the respective radiation mechanisms were dynamically coupled 
(Stamatikos et al. 2009). Furthermore, the significant excess of the prompt optical flux in comparison 
to the extrapolated $\gamma$-ray spectrum to the optical frequencies possibly indicate towards two 
different emission components for the two observed frequencies (but see Kumar \& Panaitescu 2008). 


\subsection{Multi-frequency evolution of the afterglow}

According to the standard Fireball model, the GRB afterglow is expected to be synchrotron 
radiation with the observed flux $f \propto \nu^{\beta} t^{\alpha}$, for the regions without 
spectral breaks, where the values of power-law temporal decay index ($\alpha$) and spectral index 
($\beta$) are related to each other and can be used to understand the afterglow evolution 
(Sari, Piran, Narayan 1998; Sari, 
Piran, Halpern 1999). The rich multi-frequency data of the afterglow of GRB\, 080319B available
in the literature and the data from the present work are used to measure the
indices $\alpha$ and $\beta$ of the afterglow. The values of the temporal decay indices
have been derived using empirical broken power-law relations by minimization of $\chi^{2}$ 
as described in Granot \& Sari (2002) (their equations 1 and 4). 
The bright optical flash ($10 < t < 100 ~s$) rises to 
the peak brightness ($V\sim 5.32 \pm 0.04$) with $\alpha_{V1} = 4.64 \pm 0.67$, 
and after about $50~s$ decays with $\alpha_{V2} = -4.41 \pm 0.10$.
Post $\sim 100~s$ decay of the optical afterglow could be described
as broken power law with $\alpha_{V3} = -2.33 \pm 0.23$, 
$\alpha_{V4} = -1.31 \pm 0.02$ and the break happening at 
$\sim 784.9 \pm 304.4~s$ after the burst.
The $X-$ray afterglow light curve is described by a triple broken power law
with $\alpha_{X1} = -1.40 \pm 0.01$, $\alpha_{X2} = -1.94 \pm 0.12$, 
$\alpha_{X3} = -1.14 \pm 0.09$, $\alpha_{X4} = -2.67 \pm 0.74$
and the breaks happening at $t_{b,X1} = 2583.39 \pm 871~s$,
$t_{b,X2} = (3.99 \pm 0.89)\times 10^{4}~s$ and
$t_{b,X3} = (1.0 \pm 0.29) \times 10^{6}~s$ with $\chi^{2}/dof=975/710=1.37$.
The derived values of $X-$ray temporal index $\alpha_{X4}$ and the corresponding
break time are in agreement with those derived by Tanvir et al. (2009) using late time
$X-$ray and optical data which provide evidences in favor of jet-break around 
$\sim 10^{6}~s$ after the burst.

\section{{\bf Discussion}}

\subsection{{\bf Surrounding medium}}

The closure relations between $\alpha$ and $\beta$ can also be used to
infer density profile of the circum-burst medium or to distinguish between
theoretical afterglow models like {\it ISM} and {\it WM} (Price et al. 2002; 
Starling et al. 2008). For the radiation due to a shock wave interacting with the 
{\it ISM} circum-burst medium, the expected closure relation is
$\alpha - 1.5\beta = 0.0$ in a spectral regime $\nu_{m} < \nu < \nu_{c}$, where
$\nu_{m}$ and $\nu_{c}$ are the maximum synchrotron and cooling frequencies 
, respectively. In the case of {\it WM} circum-burst medium
and for a spectral regime $\nu_{m} < \nu < \nu_{c}$ the closure relation 
is $\alpha - 1.5\beta = 0.5$. If $\nu_{m} < \nu_{c} < \nu$ then 
$\alpha - 1.5\beta = -0.5$ is expected irrespective of 
the density profile being {\it ISM} or {\it WM}.
At early times ($ t \leq 2000~s$), the optical afterglow is dominated by emission 
due to the reverse shock and at late times ($t \geq 10^{5}~s$) it is clearly dominated 
by that due to the forward shock. The values of $\beta$ estimated by Bloom et al.(2009) 
and the values of $\alpha$ of optical light curves at ($ t > 10^{5} ~s$) are consistent with the 
closure relations for both the density profiles, {\it ISM} and {\it WM} with the 
spectral regime being $\nu_{m} < \nu_{optical} < \nu_{c}$.

Also, It is clear that the temporal decay index of the optical afterglow between 
$100~s$ and $1000~s$ ($\alpha \sim -2.3$) is faster than what would be expected due 
to a forward shock scenario ($\alpha \sim -0.8$) interacting with the
circum-burst medium (Sari, Prian \& Narayan 1998). However, such a steep decay would be 
expected due a reverse shock interacting with the ejected shell. In that case,
the radiation due to interaction of the forward shock with circum-burst medium
would dominate the afterglow after the radiation from reverse shock has
died-down considerably which in this case could be happening after $10^{5}~s$.
Thus, this optical afterglow could be explained using the reverse shock
and the forward shock interactions with the circum-burst medium.

\subsection{{\bf Numerical model fits to the data}}

We have fitted the afterglow using a reverse shock (RS) model, a
forward shock (FS) model and the radiation mechanism assumed is synchrotron
following the standard Fireball scenario (Rees \& Me\'sza\'ros 1992;
Me\'sza\'ros \& Rees 1994; Kobayashi 2000). As shown in Figure (1), the multi-band
optical afterglow is very well reproduced by this model whereas $X-$ray observations
can not be reproduced by the model. Our best fit model is
consistent with no jet break in the optical afterglow and introducing a jet break
at $t\sim 10^{6}~s$ to coincide with the break in the $X-$ray afterglow does
not change the fits significantly.

The complicated behavior of the $X-$ray afterglow has been explained
by Racusin et al. (2008b) using a two-component jet model
- a central narrow jet surrounded by a co-axial wider jet.
In this model, the $X-$ray afterglow from the central narrow jet dominates
at early times, until about $2 \times 10^{4}~s$, when it fades below the brightness
of the surrounding wide jet. In this model, the breaks in the $X-$ray afterglow
light-curves at times $t_{b,X1} \sim 2\times10^{3}~s$ and
$t_{b,X3} \sim 10^{6}~s$ correspond to the jet breaks
due to the lateral spreading of narrow and wide jets, respectively.
We have tried to fit the whole data set using this two-component-jet model also,
but we found that in this model the optical radiation
from the narrow jet would be brighter than the observed optical afterglow
(See Figure 2). There is no possible way to suppress this optical radiation from
the narrow jet. The dominance of RS over FS contribution at early times of the light-curves 
are also shown for optical-$NIR$ frequencies in Figure 3 for both the models.

Alternatively, it is possible that the $X-$ray and optical afterglows
of GRB\, 080319B are not related to each other and may have independent origins.
Also, the X-ray afterglow predicted by the RS-FS model is fainter than that is 
observed and requires an additional component.

Assuming that the shock-wave is expanding into the circum-burst medium
which has a {\it WM} density profile, our best fit spectral parameters
at epoch $t = 10^{4}~s$ imply that
the peak of the synchrotron spectrum is below the optical
bands ($\nu_{m} < 1.3 \times 10^{14}$ Hz) with corresponding
normalization flux ($F_{\nu_{m}} < 607 ~\mu $Jy)
and the self-absorption frequency, $\nu_{a} > 10^{8}$ Hz. The index of electron
energy distribution {\bf $p \sim 2.07$} is lower than the canonical value $p\sim2.3$
which has normally been observed (Panaitescu \& Kumar 2001, 2002; Starling et al. 2008). 
The location of the cooling break frequency
can be constrained to be below the $X-$ray band $\nu_{c} \leq 10^{18}$ Hz
which would be consistent with the observed $X-$ray afterglow at late
times ($t > 10^{5}~s$). However, if origin of the $X-$ray afterglow is independent
of the optical afterglow then $\nu_{c} \ll 10^{18}$ Hz.

The estimated isotropic equivalent afterglow kinetic energy ($E_K^{iso}$) 
released in this explosion turns out to be 
$ < 5.5 \times 10^{53}$ erg.
For the assumed {\it WM} circum-burst medium we find 
the value of the parameter $A_{\ast} > 0.01$.
The fraction of the total energy given for accelerating electrons
($\epsilon_{e}$) and into the magnetic fields ($\epsilon_{B}$)
turn out to be about $> 0.41 $ and
$< 3 \times 10^{-3} $, respectively.
The absence of any jet break until about $10^{6}~s$ imply a jet opening
angle $> 1.4 $ degrees and hence the true amount
of released energy must be $> 8\times 10^{49}$ erg. The radius of the Fireball
at $10^{4}~s$ turns out to be $ < 5 \times 10^{18}$ cm
which we extrapolated back in time to estimate the physical size of
the emitting region during the optical flash ($t\sim50~s$) and it turns
out to be $ < 3.6 \times 10^{17}$ cm.
Similarly, the Lorentz factor of the blast-wave turns out
to be $ < 64$ at $10^{4}~s$ after
the burst which corresponds to the Lorentz factor of about
$< 240 $ at the time of the optical flash.
These findings, which are based only on the evolution of the afterglow,
are comparable with those of Racusin et al. (2008b) and Kumar \& Panaitescu
(2008) who have used arguments based on the prompt emission
to reach these conclusions.
It is reassuring that the different approaches have resulted
in similar Fireball sizes and the blast wave Lorentz factors.

The afterglow can also be explained by assuming {\it ISM} density
profile of the circum-burst medium. The best fit spectral parameters
in this case, at epoch $t = 10^{4}~s$ turn out to be
$\nu_{m} < 2.1 \times 10^{14}$ Hz with corresponding flux 
$F_{\nu_{m}} < 566 ~\mu$Jy,
and the self-absorption frequency, {\it $\nu_{a} > 10^{8}$} Hz. To explain relatively
steep decay of the afterglow the required value of $p=2.71$
turn out to be on a higher side and $\nu_{c} \leq 10^{18}$ Hz,
similar as in the case of {\it WM} density profile.

The physical parameters estimated using these best fit spectral parameters
are $E_K^{iso} < 2.4 \times 10^{52}$ erg,
$\epsilon_{e} > 0.14 $, $\epsilon_{B} < 0.02 $ and $n > 7.5\times10^{-4}$ atoms/cc.
The absence of any jet break until about $10^{6} ~s$ means
that the jet opening angle $\theta_{j} > 4.6 $ degrees and hence
the beaming corrected afterglow kinetic energy $E_{K}^{corr} > 
3.8\times10^{49}$ erg.
The radius of the Fireball at $10^{4}~s$ turns out to be
about $< 2.1 \times 10^{18}$ cm
and at the time of the optical flash $t\sim50~s$ about
$< 5.5\times10^{17}$ cm. The corresponding
Lorentz factors being $< 52.4$
and $< 382$, at $10^{4}~s$
and at $50~s$ after the burst receptively. As before, these values are also 
comparable to those of Racusin et al. (2008) and Kumar \& Panaitescu (2008).

\section{Conclusions}

Observations of the afterglow of GRB 080319B at radio, mm and optical 
frequencies are presented. The simultaneous multi-band afterglow modeling is useful 
to constrain the nature of the burst emission mechanism and the ambient medium. 
We find that the afterglow of GRB\, 080319B is consistent 
with the fireball expanding either into the {\it ISM} or into the {\it WM}
circum-burst medium. We also rule out the double-jet model as a correct 
explanation for the observed multi-band behavior of the afterglow.
While under the assumptions of our model we are able to explain the multi-band optical 
and radio afterglow reasonably well, we find that the X-ray afterglow is difficult
to explain. Alternatively, it is possible that the optical and $X-$ray afterglows might 
have different origins.
The Lorentz factor of the shock wave is estimated entirely from 
the afterglow evolution. The Lorentz factor at the time of naked eye brightness, 
extrapolated from the late estimation, turns out to be $\sim$ 300. The corresponding 
radius of the shock front is about $ 10^{17}$ cm.
We also showed that the early peak brightness of 
the afterglow could not be due to the reverse shock.
An additional emission mechanism, such as internal shocks, is required.
The results also indicate that existing blast-wave afterglow models are required 
to be modified in the light of complicated behavior of observed afterglows. 
In future, observations of prompt optical spectra will be very useful to understand 
the very early part of the afterglows of GRBs.

\begin{acknowledgements}
This research has made use of data obtained through the High Energy Astrophysics Science 
Archive Research Center On Line Service, provided by the NASA/Goddard Space Flight Center.
Partly supported by the Spanish Ministry program AYA 2007-63677. SBP acknowledge the 
Indo-Russian (DST-RFBR) project No. RUSP-836 (RFBR-08-02:91314) for this research work and
Prof. V. V. Sokolov for the useful discussions. The co-author VC, a Research Fellow, 
acknowledges the Belgian National Fund for Scientific Research (FNRS) for the financial support. 
Observation in Ussuriisk Astrophysical Observatory was made with RS-22 telescope provided by ISON.

\end{acknowledgements}

\begin{landscape}
\begin{table}
\scriptsize
\caption{Log of optical{\it-NIR} observations of the afterglow of GRB 080319B.}
\begin{tabular}{lccccccccccc} 
\hline \hline
$UT start [d]$ & $T- T_0 [d]$ & $T_{\rm exp}$ & $Filter$ &  $mag$ & $err_{mag}$ & $Telescope$\\
\hline \hline
2008 Mar. 19.4616 & 0.2027 & 1200\,s & none &  $>$15.5 &   & 0.22\,m SR-22 \\
2008 Mar. 19.7642 & 0.5052 & 600\,s & R & 19.07 & 0.19 & 0.7\,m AZT-8\\
2008 Mar. 19.7721 & 0.5132 & 600\,s & R & 19.26 & 0.27 & 0.7\,m AZT-8\\
2008 Mar. 19.7800 & 0.5211 & 600\,s & R & 19.74 & 0.50 & 0.7\,m AZT-8\\
2008 Mar. 19.7911 & 0.5322 & 600\,s & R & 19.96 & 0.39 & 0.7\,m AZT-8\\
2008 Mar. 19.7974 & 0.5385 & 600\,s & R & 19.60 & 0.34 & 0.7\,m AZT-8\\
2008 Mar. 19.8055 & 0.5466 & 900\,s & R & 19.60 & 0.20 & 0.7\,m AZT-8\\
2008 Mar. 19.8174 & 0.5585 & 900\,s & R & 19.40 & 0.15 & 0.7\,m AZT-8\\
2008 Mar. 19.8285 & 0.5696 & 900\,s & R & 19.89 & 0.28 & 0.7\,m AZT-8\\
2008 Mar. 19.8470 & 0.5881 & 900\,s & R & 19.82 & 0.23 & 0.7\,m AZT-8\\
2008 Mar. 19.8557 & 0.5968 & 900\,s & R & 20.06 & 0.19 & 0.7\,m AZT-8\\
2008 Mar. 19.8630 & 0.6041 & 900\,s & R & 19.88 & 0.23 & 0.7\,m AZT-8\\
2008 Mar. 19.8888 & 0.6299 & 720\,s & R & 20.00 & 0.27 & 0.7\,m AZT-8\\
2008 Mar. 19.8980 & 0.6391 & 720\,s & R & 19.52 & 0.16 & 0.7\,m AZT-8\\
2008 Mar. 19.9071 & 0.6482 & 720\,s & R & 20.09 & 0.28 & 0.7\,m AZT-8\\
2008 Mar. 19.9139 & 0.6550 & 720\,s & R & 19.80 & 0.24 & 0.7\,m AZT-8\\
2008 Mar. 19.9234 & 0.6645 & 900\,s & R & 19.88 & 0.27 & 0.7\,m AZT-8\\
2008 Mar. 19.9354 & 0.6765 & 900\,s & R & 20.19 & 0.40 & 0.7\,m AZT-8\\
2008 Mar. 19.9497 & 0.6908 & 900\,s & R & 19.54 & 0.15 & 0.7\,m AZT-8\\
2008 Mar. 19.9625 & 0.7036 & 900\,s & R & 20.16 & 0.44 & 0.7\,m AZT-8\\
2008 Mar. 19.9752 & 0.7163 & 720\,s & R & 20.47 & 0.25 & 0.7\,m AZT-8\\
2008 Mar. 19.9843 & 0.7254 & 960\,s & R & 20.00 & 0.37 & 0.7\,m AZT-8\\
2008 Mar. 19.9959 & 0.7370 & 900\,s & R & 20.03 & 0.37 & 0.7\,m AZT-8\\
2008 Mar. 20.0127 & 0.7538 & 720\,s & R & 20.01 & 0.22 & 0.7\,m AZT-8\\
2008 Mar. 20.0196 & 0.7607 & 720\,s & R & 19.83 & 0.21 & 0.7\,m AZT-8\\
2008 Mar. 20.0664 & 0.8075 & 720\,s & R & 19.94 & 0.23 & 0.7\,m AZT-8\\
2008 Mar. 20.0755 & 0.8166 & 720\,s & R & 20.22 & 0.26 & 0.7\,m AZT-8\\
2008 Mar. 20.0848 & 0.8259 & 900\,s & R & 20.28 & 0.24 & 0.7\,m AZT-8\\
2008 Mar. 20.0959 & 0.8370 & 900\,s & R & 19.74 & 0.16 & 0.7\,m AZT-8\\
2008 Mar. 20.1069 & 0.8480 & 600\,s & R & 19.56 & 0.20 & 0.7\,m AZT-8\\
2008 Mar. 19.8319 & 0.5730 & 600\,s & V & 19.60 & 0.70 & 0.7\,m AZT-8\\
2008 Mar. 20.0870 & 0.8351 &  1200\,s         & I & 19.57& 0.08 & 0.8\,m IAC 80 \\
2008 Mar. 19.9924 & 0.7457 &  4$\times$300\,s & R & 19.82& 0.07 & 0.8\,m IAC 80 \\
2008 Mar. 20.0557 & 0.8003 & 600\,s           & R & 19.94& 0.11 & 0.8\,m IAC 80 \\
2008 Mar. 20.1067 & 0.8548 & 2$\times$600\,s & R & 20.15& 0.08 & 0.8\,m IAC 80 \\
2008 Mar. 20.1294 & 0.8778 & 2$\times$600\,s & R & 20.14& 0.09 & 0.8\,m IAC 80 \\
2008 Mar. 20.2454 & 0.9972 & 3$\times$600\,s & R & 20.50& 0.08 & 0.8\,m IAC 80 \\
2008 Mar. 22.0730 & 2.8355 & 6$\times$600\,s & R & 22.03& 0.27 & 0.8\,m IAC 80 \\
2008 Mar. 20.0252 & 0.7814 & (1200+600)\,s & V & 20.39& 0.10 & 0.8\,m IAC 80 \\
2008 Mar. 20.0648 & 0.8165 & 3$\times$600\,s & B & 20.89& 0.12 & 0.8\,m IAC 80 \\
2008 Mar. 20.1703 & 0.9149 & 600\,s & B      & 20.75& 0.20 & 0.8\,m IAC 80 \\
2008 Mar. 19.9361 & 0.6985 & 120\,s & r & 19.56& 0.17 & 1.2\,m Mercator \\
2008 Mar. 19.9613 & 0.7031 & 120\,s & r & 19.28& 0.19 & 1.2\,m Mercator \\
2008 Mar. 19.9631 & 0.7049 & 120\,s & r & 19.74& 0.15 & 1.2\,m Mercator \\
2008 Mar. 19.9649 & 0.7067 & 120\,s & r & 19.37& 0.11 & 1.2\,m Mercator \\
2008 Mar. 19.9667 & 0.7085 & 120\,s & r & 19.93& 0.15 & 1.2\,m Mercator \\
2008 Mar. 19.9681 & 0.7095 & 60\,s & r &  19.53& 0.11 & 1.2\,m Mercator \\
2008 Mar. 19.9691 & 0.7106 & 60\,s & r &  20.07& 0.21 & 1.2\,m Mercator \\
2008 Mar. 19.9702 & 0.7117 & 60\,s & r &  19.41& 0.10 & 1.2\,m Mercator \\
2008 Mar. 19.9713 & 0.7127 & 60\,s & r &  19.66& 0.20 & 1.2\,m Mercator \\
2008 Mar. 19.9723 & 0.7138 & 60\,s & r &  19.82& 0.15 & 1.2\,m Mercator \\
2008 Mar. 19.9734 & 0.7148 & 60\,s & r &  19.68& 0.09 & 1.2\,m Mercator \\

\hline \hline
\end{tabular}
\end{table}
\end{landscape}

\begin{landscape}
\begin{table}
\scriptsize
\begin{tabular}{lccccccccccc} 
\hline \hline
$UT start [d]$ & $T- T_0 [d]$ & $T_{\rm exp}$ & $Filter$ &  $mag$ & $err_{mag}$ & $Telescope$\\
\hline \hline

2008 Mar. 19.9745 & 0.7159 & 60\,s & r &  19.61& 0.09 & 1.2\,m Mercator \\
2008 Mar. 19.9759 & 0.7177 & 120\,s & r & 19.57& 0.12 & 1.2\,m Mercator \\
2008 Mar. 19.9777 & 0.7195 & 120\,s & r & 19.54& 0.09 & 1.2\,m Mercator \\
2008 Mar. 19.9794 & 0.7212 & 120\,s & r & 19.62& 0.07 & 1.2\,m Mercator \\
2008 Mar. 19.9812 & 0.7230 & 120\,s & r & 19.83& 0.09 & 1.2\,m Mercator \\
2008 Mar. 19.9830 & 0.7248 & 120\,s & r & 19.58& 0.06 & 1.2\,m Mercator \\
2008 Mar. 19.9848 & 0.7266 & 120\,s & r & 19.65& 0.07 & 1.2\,m Mercator \\
2008 Mar. 19.9865 & 0.7283 & 120\,s & r & 19.67& 0.07 & 1.2\,m Mercator \\
2008 Mar. 19.9883 & 0.7301 & 120\,s & r & 19.78& 0.08 & 1.2\,m Mercator \\
2008 Mar. 19.9900 & 0.7318 & 120\,s & r & 19.51& 0.07 & 1.2\,m Mercator \\
2008 Mar. 19.9918 & 0.7336 & 120\,s & r & 19.66& 0.08 & 1.2\,m Mercator \\
2008 Mar. 19.9936 & 0.7354 & 120\,s & r & 19.47& 0.05 & 1.2\,m Mercator \\
2008 Mar. 19.9953 & 0.7371 & 120\,s & r & 19.62& 0.06 & 1.2\,m Mercator \\
2008 Mar. 19.9971 & 0.7389 & 120\,s & r & 19.46& 0.05 & 1.2\,m Mercator \\
2008 Mar. 19.9989 & 0.7407 & 120\,s & r & 19.62& 0.06 & 1.2\,m Mercator \\
2008 Mar. 20.1446 & 0.8871 & 120\,s & r & 19.76& 0.07 & 1.2\,m Mercator \\
2008 Mar. 20.1464 & 0.8888 & 120\,s & r & 19.95& 0.08 & 1.2\,m Mercator \\
2008 Mar. 20.1481 & 0.8906 & 120\,s & r & 20.14& 0.10 & 1.2\,m Mercator \\
2008 Mar. 20.1500 & 0.8925 & 120\,s & r & 19.78& 0.07 & 1.2\,m Mercator \\
2008 Mar. 20.1517 & 0.8942 & 120\,s & r & 20.04& 0.11 & 1.2\,m Mercator \\
2008 Mar. 20.1535 & 0.8960 & 120\,s & r & 19.98& 0.10 & 1.2\,m Mercator \\
2008 Mar. 20.1554 & 0.8979 & 120\,s & r & 19.85& 0.08 & 1.2\,m Mercator \\
2008 Mar. 20.1572 & 0.8997 & 120\,s & r & 19.85& 0.12 & 1.2\,m Mercator \\
2008 Mar. 20.1589 & 0.9014 & 120\,s & r & 19.87& 0.13 & 1.2\,m Mercator \\
2008 Mar. 20.1607 & 0.9032 & 120\,s & r & 19.87& 0.11 & 1.2\,m Mercator \\
2008 Mar. 20.1625 & 0.9049 & 120\,s & r & 19.92& 0.09 & 1.2\,m Mercator \\
2008 Mar. 20.1642 & 0.9067 & 120\,s & r & 19.96& 0.11 & 1.2\,m Mercator \\
2008 Mar. 20.1660 & 0.9085 & 120\,s & r & 20.04& 0.10 & 1.2\,m Mercator \\
2008 Mar. 20.1678 & 0.9103 & 120\,s & r & 19.84& 0.08 & 1.2\,m Mercator \\
2008 Mar. 20.1695 & 0.9120 & 120\,s & r & 19.73& 0.09 & 1.2\,m Mercator \\
2008 Mar. 20.1713 & 0.9138 & 120\,s & r & 19.84& 0.11 & 1.2\,m Mercator \\
2008 Mar. 20.1731 & 0.9156 & 120\,s & r & 19.92& 0.10 & 1.2\,m Mercator \\
2008 Mar. 20.1748 & 0.9173 & 120\,s & r & 20.01& 0.12 & 1.2\,m Mercator \\
2008 Mar. 20.1766 & 0.9191 & 120\,s & r & 19.91& 0.10 & 1.2\,m Mercator \\
2008 Mar. 20.1784 & 0.9208 & 120\,s & r & 20.08& 0.09 & 1.2\,m Mercator \\
2008 Mar. 20.1801 & 0.9226 & 120\,s & r & 19.71& 0.07 & 1.2\,m Mercator \\
2008 Mar. 20.1819 & 0.9244 & 120\,s & r & 19.87& 0.09 & 1.2\,m Mercator \\
2008 Mar. 20.1837 & 0.9261 & 120\,s & r & 19.83& 0.07 & 1.2\,m Mercator \\
2008 Mar. 20.1854 & 0.9279 & 120\,s & r & 19.84& 0.09 & 1.2\,m Mercator \\
2008 Mar. 21.8618  & 2.6029 & 5760\,s & I &  21.22 &0.15 & 1.5\,m AZT-22  \\
2008 Mar. 21.8781  & 2.6192 & 5610\,s & R &  21.81 &0.26 & 1.5\,m AZT-22  \\
2008 Apr. 07.0086  & 18.7479 & 7200\,s & R &  $>$23.6 & & 1.5\,m   AZT-22  \\
2008 Mar. 19.9179  & 0.6590 & 3$\times$600\,s & R & 19.66& 0.09 & 2.0\,m IGO \\
2008 Mar. 27.9511  & 8.6524 & 28$\times$120,s & R &  $>$23.0 & & 2.5\,m   NOT  \\
\hline
\end{tabular}
\end{table}
\end{landscape}

\end{document}